# Demonstration and STEM Analysis of Ferroelectric Switching in MOCVD-Grown Single Crystalline Al0.85Sc0.15N

Niklas Wolff*, Georg Schönweger*, Isabel Streicher, Md Redwanul Islam, Nils Braun, Patrik Straňák, Lutz Kirste, Mario Prescher, Andriy Lotnyk, Hermann Kohlstedt, Stefano Leone[a], Lorenz Kienle[b], Simon Fichtner[c]

*Authors contributed equally to this work.

Dr. N. Wolff (niwo@tf.uni-kiel.de, orcid-org/0000-0002-8796-0607), Md R.Islam (mdis@tf.uni-kiel.de), Prof. Dr. L.Kienle ([b]lk@tf.uni-kiel.de), Dr. S. Fichtner ([c]sif@tfuni-kiel.de)
Department of Material Science, Kiel University, Kaiserstrasse 2, D-24143 Kiel, Germany

G. Schönweger (gmsc@tf.uni-kiel.de), Prof. Dr. H. Kohlstedt (hko@tf.uni-kiel.de)
Department of Electrical and Information Engineering, Kiel University, Kaiserstrasse 2, D-24143 Kiel, Germany

G. Schönweger, Dr. S. Fichtner
Fraunhofer Institute for Silicon Technology (ISIT), Fraunhoferstr. 1, D-25524 Itzehoe, Germany

N. Braun (nils.braun@iom-leipzig.de), Dr. A. Lotnyk (andriy.lotnyk@iom-leipzig.de)
Leibniz Institute of Surface Engineering (IOM), Permoserstr. 15, D-04318 Leipzig, Germany

I. Streicher (isabel.streicher@iaf.fraunhofer.de), P. Straňák (patrik.stranak@iaf.fraunhofer.de), M. Prescher (mario.prescher@iaf.fraunhofer.de) Dr. L. Kirste (Lutz.Kirste@iaf.fraunhofer.de), Dr. S. Leone ([a]stefano.leone@iaf.fraunhofer.de) Fraunhofer Institute for Applied Solid State Physics (IAF), Tullastrasse 72, D-79108 Freiburg, Germany

Dr. N. Wolff, Prof. Dr. L.Kienle, Dr. S. Fichtner, Prof. Dr. H. Kohlstedt
Kiel Nano, Surface and Interface Science (KiNSIS), Kiel University, Christian-Albrechts-Platz 4, D-24118 Kiel, Germany

Keywords: wurtzite ferroelectrics, III-nitrides, AlScN, STEM, nanostructure

## Abstract

Wurtzite-type Al1-xScxN solid solutions grown by metal organic chemical vapour deposition are for the first time confirmed to be ferroelectric. The film with 230 nm thickness and x = 0.15 exhibits a coercive field of 5.5 MV/cm at a measurement frequency of 1.5 kHz. Single crystal quality and homogeneous chemical composition of the film was confirmed by X-ray diffraction spectroscopic methods such as time of flight secondary ion mass spectrometry. Annular bright field scanning transmission electron microscopy served to proof the ferroelectric polarization inversion on unit cell level. The single crystal quality further allowed to image the large-scale domain pattern of a wurtzite-type ferroelectric for the first time, revealing a predominantly cone-like domain shape along the c-axis of the material. As in previous work, this again implies the presence of strong polarization discontinuities along this crystallographic axis, which could be suitable for current transport. The domains are separated by narrow domain walls, for which an upper thickness limit of 3 nm was deduced, but which could potentially be atomically sharp. We are confident that these results will advance the commencing integration of wurtzite-type ferroelectrics to GaN as well as generally III-N based heterostructures and devices.

# 1 Introduction

Solid solutions with wurtzite-type crystal structure are one of the newest material class to have been confirmed as possessing ferroelectric properties. [1, 2] They are characterized by some of the largest spontaneous polarizations and coercive fields among ferroelectrics in general, extreme temperature stability as well as compatibility to the major Si and GaN based semiconductor platforms.[3, 4, 5, 6, 7] The fact that the wurtzite-type ferroelectrics share the same crystal structure and are most often III-nitrides themselves, offers the perspective of introducing novel heterostructures with a sort of native ferroelectricity to engineer novel GaN based devices. One of the first devices considered in this context is the ferroelectric high electron mobility transistor (FE-HEMT), [5, 6, 7, 8] which could ultimately allow reconfigurable normally-off functionality in normally-on devices. Due to economic benefits like high yield and ease-of-operation as well as very good single crystal quality, metal organic chemical vapor deposition (MOCVD) is today the dominant method for III-nitride-based HEMT fabrication. The availability of MOCVD-grown wurtzite-type ferroelectric single crystal $Al_{1-x}Sc_xN$ thin films would therefore be of great benefit to advance the integration of said materials to GaN technology. However, although the MOCVD growth of $Al_{1-x}Sc_xN$ has already been demonstrated in the last years, targeting either HEMT or RF-filters, [9, 10, 11, 12] no demonstration of ferroelectricity has been reported so far. In this work, we demonstrate for the first time that single crystalline $Al_{0.85}Sc_{0.15}N$ films grown on GaN templates by MOCVD are indeed ferroelectric. Asides from standard electrical characterization to deduce the displacement current originating from the polarization inversion and to confirm a particularly high coercive field of 5.5 MV/cm, we focus on high resolution scanning transmission electron microscopy (STEM) for the imaging of the atomic structure. Annular bright field (ABF)-STEM is employed not only to confirm the electric field induced polarization inversion on unit cell level, but also to reveal the formation of cone-like shaped domain patterns, which we think could be the dominant domain shape in all wurtzite-type ferroelectrics in general.

# 2 Results & Discussion

MOCVD allows the deposition of $Al_{1-x}Sc_xN$ thin films with high rates and single crystalline quality – unlike sputtering and molecular beam epitaxy (MBE).[13, 14] For this study, a 230 nm thick $Al_{0.85}Sc_{0.15}N$ film was grown onto a Si-doped GaN layer deposited on a c-plane sapphire substrate at a temperature of 1000 °C using a novel precursor molecule with increased vapor pressure as described in the experimental details. As the high growth temperatures can potentially result in phase instability of the wurtzite-type metastable $Al_{1-x}Sc_xN$ solid solutions, [15, 16] the crystal structure and chemical composition of the film has been characterized by high resolution x-ray diffraction (HRXRD), ToF-SIMS and TEM methods. XRD analysis from the single crystalline film exhibit $Al_{1-x}Sc_xN$ c-axis texture with narrow rocking curve full-width at half maximum of 252 arcsec (Supporting Figure A1) and showed no secondary reflections from deviating crystal orientations or phases. The ToF-SIMS depth profile of the 230 nm thick $Al_{0.85}Sc_{0.15}N$ layer deposited on the n-GaN buffer is shown in Supporting Figure A2. No fluctuations of the chemical composition along the growth direction were observed. The intensity ratio of the AlCs+ and ScCs+ signals obtained in positive-ion detection mode were used to determine the Sc concentration to be x = 0.13, which is in good agreement with the determined Sc content of x = 0.15 via EDS (see Section 4). Both 12C- and 18O- signal intensities of the $Al_{1-x}Sc_xN$ are elevated with respect to that of the n-GaN layer. Carbon can be introduced as residual impurity in the reactor by the hydrocarbon-based precursor molecule and act as electron trap. Sc has a high oxygen affinity compared to that of

Ga and residual oxygen impurities in the precursor (research grade) and/or in the reactor can lead to a higher oxygen incorporation. In addition to the XRD analysis which, e.g., could be unable to localize or detect coherent nanoscale cubic domains, the present film was examined for local chemical variations with energy-dispersive X-ray spectroscopy (EDS) and local structural irregularities using scanning transmission electron microscopy. Selected area electron diffraction (SAED) experiments covered the complete film thickness by selecting an aperture corresponding to an investigated circular area of 250 nm in diameter. The elemental maps (Supporting Figure A3a) of the Pt/SiN/Al$_{1-x}$Sc$_x$N/GaN heterostructure show no variation of the Sc concentration along film growth direction and the average Sc content was determined as x = 0.15 by multiple EDS measurements. In agreement with XRD data, the SAED experiments do not indicate structural variation (Supporting Figure A3b) in the probed sample volume confirming the single-crystalline structure of the film.

## 2.1 Electrical Characterization

The ferroelectric nature of the heterostructure is unveiled via current density over electric field (J −E) measurements. Figure 1 depicts the ferroelectric current response of a Pt/SiN(10 nm)/Al$_{0.85}$Sc$_{0.15}$N(230 nm)/GaN capacitor with 20 µm diameter. The in situ deposited SiN acts as protection layer to prevent the Al$_{0.85}$Sc$_{0.15}$N layer from oxidation during handling. The hysteretic current contributions typical of the wurtzite-type ferroelectrics are clearly discernible at high electric fields with superimposed leakage contributions. [17, 18] The hysteretic behavior of the switching event is further confirmed via recording non-hysteretic currents by applying two consecutive unipolar voltage sweeps with the same polarity (Figure 1 - black curves). Additionally, the coercive field as well as the hysteretic area increases with increasing frequency (see Supporting Figure A4), as it is expected for ferroelectricity. The coercive field ($E_c \approx 5.5$ MV/cm at 1.5 kHz) and the non-switching leakage currents are in the range of what is typically reported for sputter-deposited as well as for MBE grown ferroelectric Al$_{0.85}$Sc$_{0.15}$N thin films.[18, 17] These findings lead to the conclusion that the deposition method and the associated crystal quality (which for MOCVD grown films is significantly higher than for sputter-deposited films - see structural analysis below) is not the major factor defining the coercive field and leakage in Al$_{1-x}$Sc$_x$N solid solutions.

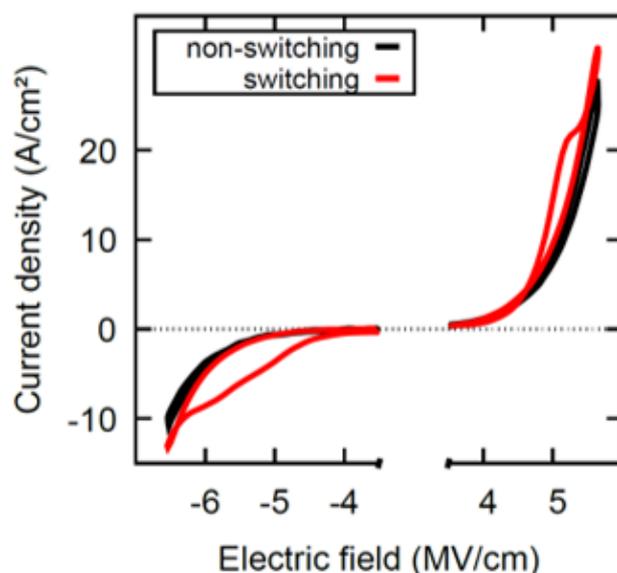

Figure 1: Current response of switching- and non-switching cycles measured at 1.5 kHz. The unipolar non-switching response was measured by pre-poling the capacitor to the respective polarity.

This may be due to the fact that sputter-deposited films consist of columnar grains with diameters in the range of 5 to 30 nm, which are highly c-axis oriented. [19]. As shown in Section 2.2, the horizontal grain size is in the same order of magnitude as the horizontal width (10-20 nm) of the domains imaged in the single crystal MOCVD film of this study. The coinciding length scale of typical sputter deposited grains and the ferroelectrically induced polarization domains can explain why to date no fundamental differences in the ferroelectric response of epitaxial and non-epitaxial wurtzite-type ferroelectrics were observed. Thus, the absence of grain boundaries in the MOCVD grown films (see Section 2.2) does not significantly impact the ferroelectric behavior and grain boundaries do not dominate the leakage mechanism of $Al_{1-x}Sc_xN$.

However, for this MOCVD-grown film, huge hysteretic currents at positive fields are observed when measuring as-deposited capacitors, while for negative fields the typical ferroelectric displacement current peak is present from the very first cycle (see Supporting Figure A5). In one of our recent studies we reported on hysteretic current contributions likely arising due to domain wall formation and conduction during polarization inversion. [20] Also in this MOCVD-grown film, polarization domain walls with horizontal components are evidenced in the cone-like shape domain patterns by STEM analysis (see Figure 4). However, in contrast to tail-to-tail domain walls (i.e., polarization direction of the adjacent domains pointing apart, resulting in a negatively charged domain wall) present in 5 nm thin films reported in our previous study, [20] in this heterostructure head-to-head domain walls (i.e., polarization direction of the adjacent domains pointing towards each other, resulting in a positively charged domain wall) are forming during the switching process. Such head-to-head domain-walls typically exhibit a much higher conductivity as they can be compensated by electrons. [21, 22] This could explain the huge hysteretic leakage current contributions observed during the first cycles. As visible in Supporting Figure A5, for positive fields the hysteretic area decreases significantly with increasing number of switching cycles and the for $Al_{1-x}Sc_xN$ typical ferroelectric displacement current peaks appear also on the positive branch after cycling ≈ 1000 times. Positive-up negative-down (PUND) measurements of such a cycled sample, which correct for non-hysteretic leakage (see Supporting Figure A6), suggest a remnant polarization of $P_{r,neg}$ = 154 $\mu C/cm^2$ and $P_{r,pos}$ = 176 $\mu C/cm^2$ for the negative and positive branches, respectively. Thus, the averaged $P_r$ of this MOCVD-grown $Al_{0.85}Sc_{0.15}N$ ($P_r$ = ($P_{r,neg}+P_{r,pos}$)/2 = 165) is in the range of the typically reported $P_r$ values of sputter-deposited as well as MBE-grown $Al_{1-x}Sc_xN$ films with similar Sc contents ($P_r$ ≈ 150 $\mu/cm^2$).[23, 24] However, the asymmetry suggests the presence of hysteretic leakage contributions likely arising due to domain wall conduction upon ferroelectric switching, as discussed above. This can serve as an explanation as to why the measured remnant polarization exceeds theoretical expectations. Similarly, to epitaxially grown $Al_{1-x}Sc_xN$ on M-polar GaN templates via MBE or via sputter-deposition, the polarity of the as-deposited state of the MOCVD-grown $Al_{0.85}Sc_{0.15}N$ is M-polar, as evidenced by atomic resolution ABF-STEM (see Figure 3) and via the electrical measurements presented in Figure 2. [23, 17] When applying a unipolar negative voltage sweep to an as-deposited capacitor (Figure 2a, red curve), no hysteretic current response and thus no ferroelectric switching of N-polar volume occurs. A subsequent negative voltage sweep results in exact the same current response, confirming the absence of as-deposited N-polar volume (black curve). Contrarily, as depicted in Figure 2b, when consecutively sweeping positive voltages on a pristine as-deposited capacitor, hysteretic current responses appear. The ferroelectric switching of as-deposited M-polar volume to N-polarity at positive voltage sweeps is confirmed by recording the current response of a subsequent negative voltage sweep on the same capacitor: A typical ferroelectric displacement peak appears, evidencing the prior inversion of the full volume from the as-deposited M-polarity into N-polarity (Figure 2c). In contrast to the typically N-polar growth of sputter-deposited $Al_{1-x}Sc_xN$ on non-polar templates, [1] retaining the polarity of the polar template lowers the polarization discontinuity and the associated energy at the interface. This driving force at the interface favoring M-polarity in $Al_{1-x}Sc_xN$ is likely the reason for the formation of the head-to-head domains present in this heterostructure. Due to that, when switching from full M-polarity to Npolarity, it is energetically more favourable to nucleate N-polar inversion domains at the top-interface (Pt/SiN), with subsequent domain growth towards the bottom interface (GaN). If this polarization reversal is not fully saturated, this results in N-polarity in the upper region of the film followed by M-polar domains in the lower region of the film, which corresponds to the head-to-head domain configuration confirmed by ABF-STEM experiments. This is in strong contrast to the aforementioned sputtered films grown on metal electrodes, with no driving force for M-polarity from the bottom interface. This results in N-polar residues at the latter when transitioning from M-polarity. [25, 20] As a consequence only tail-to-tail domain walls were observed.

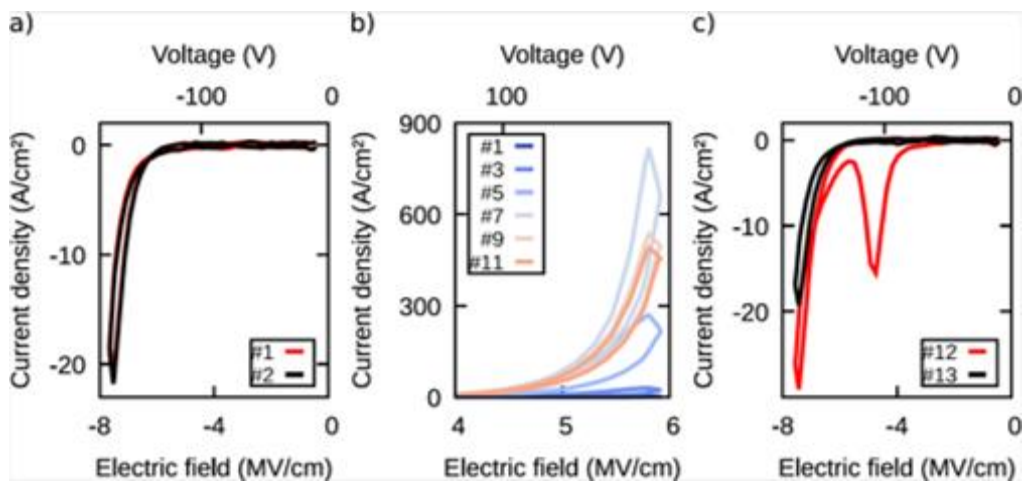

Figure 2: a) Two consecutive unipolar measurements performed on a pristine capacitor for negative electric fields revealing no as-deposited N-polarity. b) First eleven consecutive unipolar measurements performed on a pristine capacitor for positive electric fields. Hysteretic currents up to the 9th voltage pulse are visible, indicating as-deposited M-polarity which partially switches to N-polarity. c) Unipolar measurements for negative fields performed subsequently to the measurement described in b). Clearly, a polarization switch from N-polarity to M-polarity is visible, which in turn gives evidence of prior switching from as-deposited M- to N-polarity. All measurements were performed at 1.5 kHz on capacitors with 20 µm diameter.

## 2.2 Structural Characterization

As demonstrated in our previous works on sputtered $Al_{1-x}Sc_xN$ thin films, unambiguous evidence of ferroelectric polarization inversion can be reliably provided by the contrast produced by ABF-STEM. [25, 17, 20] This can be achieved by imaging the resulting unit cell inversion along the c-axis of the film. To understand the evolution of the atomic structure of single crystalline MOCVD-grown $Al_{0.85}Sc_{0.15}N$ thin films subjected to large electric fields during cycling, both the as-deposited and the switched state are compared. Unlike in our previous STEM investigations, the superior film quality enables the direct interpretation of the atomic polarization direction over large areas by the ABF-STEM image contrast examined in the as-deposited capacitor structure shown in Figure 3a. The center ABF-STEM micrograph shows the Pt/SiN/Al0.85Sc0.15N/GaN epitaxial heterostructure and the absence of structural defects, which would appear with darker contrast due to electron scattering and changes of the diffraction conditions. The atomic polarization of the as-deposited Al0.85Sc0.15N film is determined to be M-polar growing from the GaN interface to the top of the film, consistent with the electrical data. Additionally, ferroelectric switching on unit cell level of a pre-cycled (400x) capacitor was investigated as well. Here, N-polarity is expected due to the prior application of four consecutive unipolar positive voltage signals with E > Ec. An ABF-STEM image of such a switched film is presented in Figure 3b, showing the appearance of cone-like contrast features within the switched part of the thin film which are absent in the pristine film.

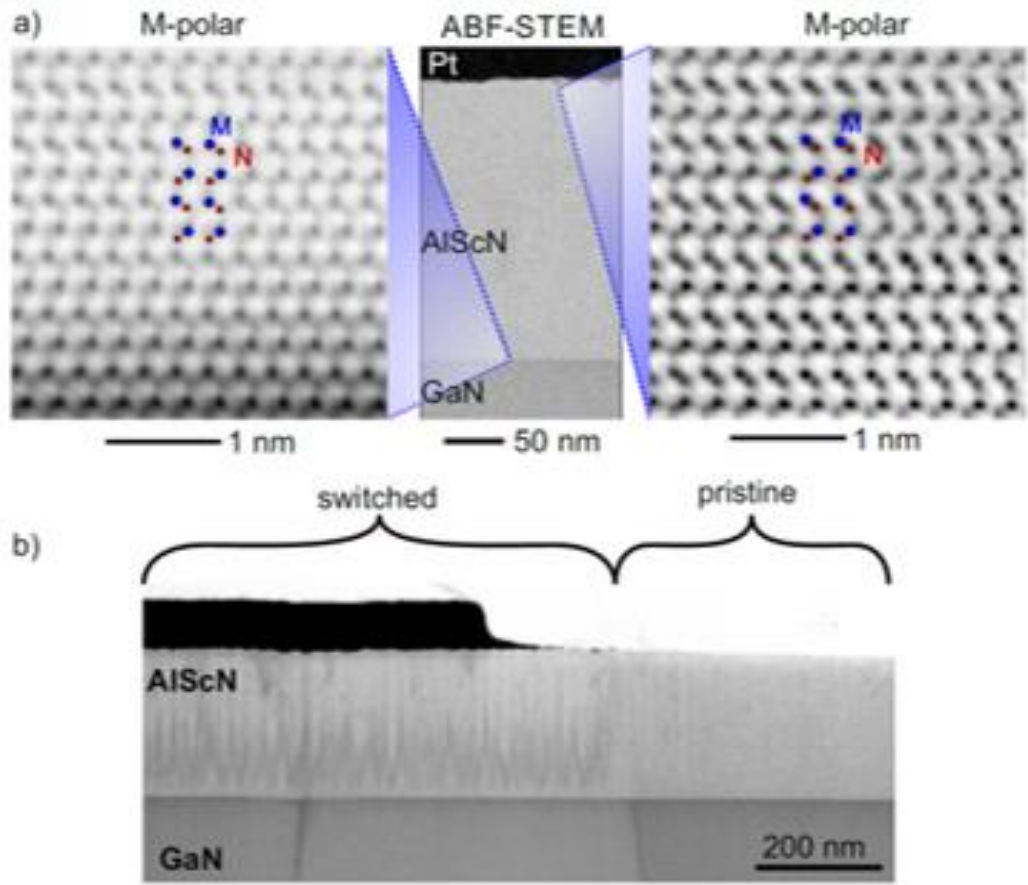

Figure 3: a) ABF-STEM investigation of the pristine MOCVD-grown Al0.85Sc0.15N layer (center image). The atomic polarization of the pristine film under the capacitor is identified to grow completely M-polar from the bottom interface (left image) towards the top electrode (right image). b) ABF-STEM image showing the cone-like domain pattern in the switched region of the film in direct comparison to the pristine region.

Further STEM investigation of the switched film presented in Figure 4 highlights the appearance of domains with N-polarity extending from the top Pt/SiN interface into the bulk of the film, providing direct confirmation for ferroelectricity in the MOCVD-grown Al0.85Sc0.15N film. In more detail, the single crystal quality, which is not disturbed by visible grain boundaries allowed the direct observation of electric-field induced structural modifications, e.g., inversion domains and their boundaries. The appearance of these structures are observed in the overview ABF-STEM micrograph (Figure 3b and Figure 4a, center) showing the formation of a zigzag and cone-like shaped contrast patterns in the bottom third of the film facing the GaN interface. The contrast patterns start to extend in approximately 30 - 50 nm distance from the

GaN interface, are 10-20 nm wide and reach up to the center of the film. Atomic resolution ABF-STEM micrographs recorded in the top 50 nm show the full inversion of the atomic structure to N-polarity (Figure 4a, right). In contrast, atomic resolution ABF-STEM micrographs recorded at the GaN interface (Figure 4a, left) and within this 30 - 50 nm transition region show remaining domains with M-polarity. The formation of this domain pattern is expected from our observation of domain patterns in 5 nm $Al_{1-x}Sc_xN$ films with horizontal (cone-like) contributions. [20] The nucleation mechanisms of the N-polar domain starts at the top electrode and subsequently progresses into the bulk of the film, forming cone-like shapes towards the pinned M-polar domain at the substrate interface. The boundaries between M- and N-polar regions were studied in more detail by ABF-STEM. This revealed a discontinuity of the M-polar domain displayed on the right side of Figure 4b, which changed incrementally across 4 unit cells to the N-polar configuration observed on the left side of the micrograph.

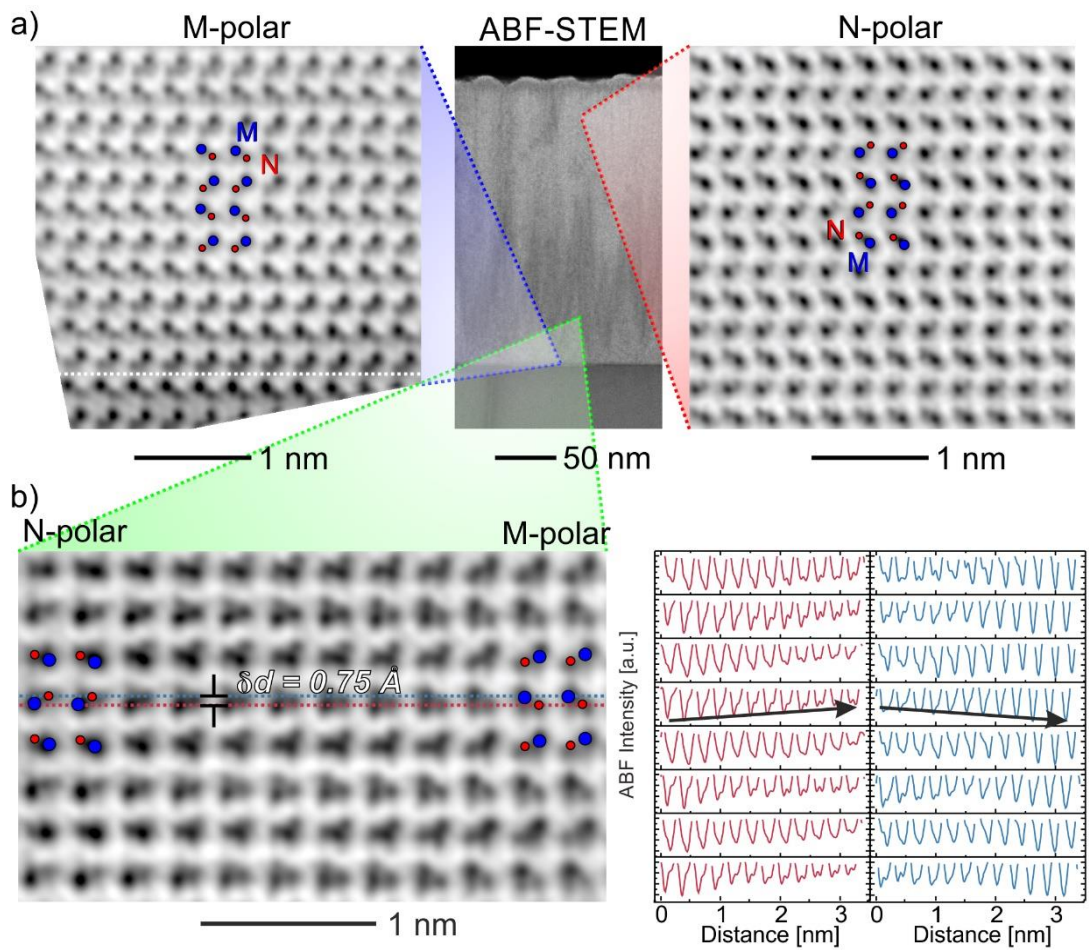

Figure 4: ABF-STEM investigation of the MOCVD-grown $Al_{0.85}Sc_{0.15}N$ layer switched to N-polarity. a) Atomic resolution micrographs showing remaining M-polar domains at the GaN interface (dotted white line) and the inversion of the unit cell to N-polarity in areas towards the top electrode. Distinct structural features are observed in the ABF-STEM image assigned to the introduction of domain walls between domains with opposite polarity. b) ABF-STEM image of a horizontal discontinuity of the M-polar domain changing to the N-polar domain across a span of 4 unit cells. Intensity profiles drawn across this transition region from left to right are centered on the metal atom position of the of N- and M-polar unit cells in each row of atoms.

In this transition region, the N-polar double spot pattern changes to a characteristic triplet of spots by the superposition with the M-polar double spot pattern. The ABF intensity across the transition region is evaluated from profiles centered on the atomic positions of the metal atoms of N- (red line) and M-polar (blue line) unit cells. Apparently, the metal atom position of the N-polar unit cell is horizontally leveled with the Nitrogen atom position of the M-polar unit cell, meaning the two atomic planes switched their position along the z direction. This corresponds to a shift of the metal atom position by Δd ≈ 0.75 Ångstrom and is consistent with experimental data on face-to-face inversion domain boundaries in GaN.[26] However, in this observation, no direct face-to-face boundary but subtle changes of ABF intensity between both domains are displayed within the individual spots of the triple spot pattern. The decrease in intensity of the N-polar metal atom position (red line) suggests the reduction of atom density in the probed metal atom column (first spot) and the change to the Nitrogen atom column of the M-polar unit cell (second spot). The same consideration holds true for the blue profiles starting on the N-polar Nitrogen atom column which intensity decreases on favor of the metal atom position belonging to the M-polar domain in the upper left spot of the triplet. Hence, the view onto an inclined inversion domain boundary (i.e., two overlapping wedge like domain volumes) leading to the observed intensity distributions is suggested. For rationalization of this assumption, a structure model containing a regular inclined inversion domain boundary was constructed. First, the wurtzite-type unit cell was transformed into an orthohexagonal (P1, 5a 6b c) unit cell to create a model with a transition region spanning four unit cells. The M-polar cell was constructed from the N-polar cell by inversion of the axes b and c. Next a (110)-inclined boundary was constructed by systematically deleting unit cells from the supercell as shown in Supporting Figure A7. Next, both cells were combined and simulations of the ABF-contrast were conducted along the b-c projection orientation presented in Figure 5. The overlap of atoms in the projection direction resembles a structure motif consisting of nitrogen and metal atom dumbbells across the constructed transition region. A similar structural motif has been reported in context of a proposed intermediate and non-polar structure occurring during polarization switching in wurtzite-type $Al_{1-x}B_xN$.[27] Although the simulated contrast distribution does not perfectly resembles the more pronounced intensity within the experimentally observed triple spot structure, the intensity profiles do provide a decent match regarding the trend of the intensity distribution across the transition region. Hence, the inclined superposition of the two domain states is currently the most likely explanation, especially considering that no stable non-polar state has been predicted for $Al_{1-x}Sc_xN$.[28] Still, more experimental work has to be carried out in the future to understand the atomic structures of the inversion domain boundaries in wurtzite-type ferroelectric materials such as $Al_{1-x}Sc_xN$ and their propagation during the switching process.

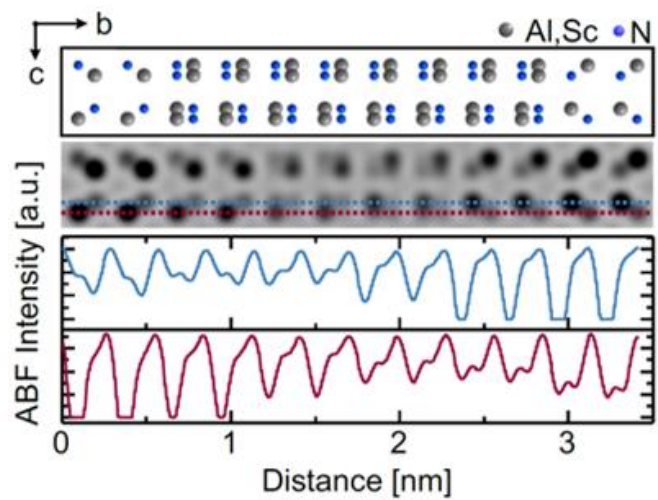

Figure 5: Supercell structure model of a simple inclined inversion domain boundary viewed along the b-c plane ([2⁻ 1 ⁻ 10] zone axis) and the corresponding simulated image and intensity profiles.

## 3 Conclusion

Ferroelectric operation of single crystal quality 230 nm thin MOCVD-grown Al0.85Sc0.15N deposited onto a GaN template was demonstrated. Electrical J−E measurements reveal a high coercive field of Ec ≈ 5.5 MV/cm at 1.5 kHz and PUND measurements suggest a remanent polarization of Pr = 165µC/cm2. The as-grown polarity of the Al0.85Sc0.15N layer was determined to be fully M-polar. Furthermore, during the first switching cycles, large hysteretic currents are contributing to the ferroelectric displacement currents at positive fields. Both effects are most likely related to the growth onto the M-polar GaN template. Retaining M-polarity in Al0.85Sc0.15N minimizes the polarization discontinuity at the interface. Contrarily, the introduction of polarization domain boundaries with horizontal component, as evidenced via STEM, implies the presence of strong polarization discontinuities in Al0.85Sc0.15N along the crystallographic c-axis, which could be suitable for current transport. Due to the driving force for M-polarity at the interface to the GaN, this discontinuity is in a head-to-head like configuration with typically associated enhanced conductivity. Therefore, domain-wall formation and conduction during the first switching cycles could explain the observed high hysteretic currents. ABF-STEM investigations of as-deposited and switched thin films resolved the evolution of the nanostructure from a single M-polar domain state into M-polar and N-polar domains after ferroelectric cycling of the material. The single crystal quality of the grown film further allowed to image the large-scale domain pattern for the first time, due to the absence of grain boundaries. The change of ABF-contrast after switching revealed a predominantly cone-like domain shape along the c-axis of the material which is consistent with our previous observations of domain boundaries with horizontal components in 5 nm thin Al1-xScxN films. These cone-like structures were identified as 2-3 nm wide boundaries between original M-polar and switched N-polar regions via resolving the atomic structure change by ABF-STEM and supercell simulation. In projection direction, the visible triple spot structure and the progressing change in contrast intensity across the individual spots within the boundary region was explained by simulation of an inclined superposition structure. These films are ideal candidates for future experiments targeting the atomic structure of the inversion domain boundaries in more detail and in situ switching experiments. As an outlook, the demonstration of ferroelectricity in Al1-xScxN thin films grown by MOCVD will advance the commencing integration of wurtzite-type ferroelectrics to GaN as well as generally III-N based heterostructures and devices.

## 4 Experimental Section

A close-coupled showerhead MOCVD reactor equipped with a proprietary setup for the generation of an adequate molar flow of the low-vapor pressure scandium precursor was used for growth.[29] A 230 nm – thick Al0.85Sc0.15N layer was grown on a 2 µm - thick Si-doped GaN layer deposited on a c-plane sapphire template. The new Sc precursor (EtCp)3Sc(bdma) provided by Dockweiler Chemicals GmbH that, thanks to its increased vapour pressure, allows for increased growth rates was employed and heated to 100 °C. Ammonia (NH3) was used as N source, trimethylaluminum (TMAI) as the group 13- or IIIA-precursor. The carrier gas was hydrogen. A 10 nm - thick SiNx in situ passivation layer was deposited on top of the Al0.85Sc0.15N to protect the layer from oxidation during ex situ handling. The growth temperature was 1000 °C and the Sc/Al flow ratio was 50, similar to our descriptions in previous reports.[14, 9] The chemical composition of the layers was characterized by time-of-flight secondary ion mass spectrometry (ToF-SIMS). The ToF-SIMS depth profile of the MOCVD grown heterostructure was obtained in a dual beam mode using a sputter beam of 1 keV Cs+

ions together with a primary beam of 30 keV 1Bi+ ions for the crater analysis. The Sc concentration was evaluated by relating the AlCs+/ScCs+ signal ratio to that of magnetron-sputtered AlScN reference samples, previously calibrated by energy elastic recoil detection analysis (ERDA).[15] The average concentration of Scandium in the AlScN layer was determined to be x = 13.4. 100 nm thick Pt was deposited on top of the heterostructure via magnetron sputtering (Von Ardenne CS 730S) and top-electrodes where structured via lithography and ion beam etching. Electrical measurements were performed using an AixACCT TF Analyzer 2000. X-ray diffraction analysis of the sample was performed in a Rigaku Smartlab SE instrument (9kW, Cu-K$\alpha$, Hypix-3000 pixel detector) equipped with a 2-bounce Ge(220) monochromator at the incident side. Cross-section samples were cut and extracted from the Pt/SiN/Al$_{0.85}$Sc$_{0.15}$N/GaN capacitor structures orthogonal to the $2\bar{1}\bar{1}0$ orientation and thinned to electron transparency by the focused ion-beam (FIB) method. Final polishing with 50 eV Ar+-ions was performed using a Model1040 NanoMill system (Fischione Instruments). Atomic scale investigation of the local atomic polarization were conducted on a probe CS-corrected JEOL (ARM200F) NEOARM scanning transmission electron microscope (STEM) operated at the acceleration voltage of 200 kV. The annular bright-field (ABF)-STEM imaging mode was chosen for the identification of low-angle scattering nitrogen atom positions in the unit cell using a collection angle-range of 10 - 20 mrad. Scan distortions and sample drift during image acquisition were minimized by fast serial recording of multi-frame images followed by post-processing image alignment. The rigid and non-rigid image registration of typically 12 images recorded at a scan speed of 12 µs/row was performed using the Smart Align algorithm [30] (HREM Research Inc.) on the DigitalMicrograph v.3.5.1 (DM) (GatanInc) software. Fourier-filtering of non-rigidly processed ABF-STEM micrographs was applied using a simple radiance difference filter (lite version of DM plug-in HREM-Filters Pro/Lite v.4.2.1, HREM Research Inc.) to remove high-frequency noise from the post-processed image. ABF image simulations were performed using the Dr.Probe software [31] and idealized microscope parameters (semicoherent probe, probe forming aperture 35 mrad) setting all aberrations to zero. The film stoichiometry was determined by energy-dispersive X-ray spectroscopy (EDS) using a dual silicon drift detector system with 100 mm² active area each. The denoted scandium content x is defined as the number of Sc atoms relative to the total number of metal atoms (Sc + Al) with an estimated uncertainty of ±0.01.

**Acknowledgements**

This collaborative work is enabled through funding by the Federal Ministry of Education and Research (BMBF) in projects "ForMikro-SALSA" (Project-ID 16ES1053) and ProMat_KMU "PuSH" Grant Number 03XP0387B and the Deutsche Forschungsgemeinschaft (DFG, German Research Foundation) – Project-ID 434434223 – SFB 1461; Project-ID 286471992 - SFB 1261 as well as Project-ID 458372836 and Project-ID 448667535. We would like to thank Lars Thormählen for the deposition of the Pt layer.

**References**

[1] S. Fichtner, N. Wolff, F. Lofink, L. Kienle, B. Wagner, Journal of Applied Physics 2019, 125, 11 114103.

[2] P. Wang, D. Wang, S. Mondal, M. Hu, J. Liu, Z. Mi, Semiconductor Science and Technology 2023, 38, 4043002.

[3] M. R. Islam, N. Wolff, M. Yassine, G. Schönweger, B. Christian, H. Kohlstedt, O. Ambacher, F. Lofink, L. Kienle, S. Fichtner, Applied Physics Letters 2021, 118, 23 232905.

[4] S. Fichtner, G. Schönweger, F. Dietz, H. Hanssen, H. Züge, T.-N. Kreutzer, F. Lofink, H. Kohlstedt, H. Kapels, M. Mensing, In 2023 7th IEEE Electron Devices Technology Manufacturing Conference (EDTM). 2023 1–3.

[5] J. Casamento, K. Nomoto, T. S. Nguyen, H. Lee, C. Savant, L. Li, A. Hickman, T. Maeda, J. Encomendero, V. Gund, A. Lal, J. C. M. Hwang, H. G. Xing, D. Jena, In 2022 International Electron Devices Meeting (IEDM). 2022 11.1.1–11.1.4.

[6] J. Y. Yang, S. Y. Oh, M. J. Yeom, S. Kim, G. Lee, K. Lee, S. Kim, G. Yoo, IEEE Electron Device Letters 2023, 1–1.

[7] Z. Zhao, Y. Dai, F. Meng, L. Chen, K. Liu, T. Luo, Z. Yu, Q. Wang, Z. Yang, J. Zhang, W. Guo, L. Wu, J. Ye, Applied Physics Express 2023, 16, 3 031002.

[8] D. Wang, P. Wang, M. He, J. Liu, S. Mondal, M. Hu, D. Wang, Y. Wu, T. Ma, Z. Mi, Applied Physics Letters 2023, 122, 9 090601.

[9] I. Streicher, S. Leone, C. Manz, L. Kirste, M. Prescher, P. Waltereit, M. Mikulla, R. Quay, O. Ambacher, Crystal Growth & Design 2023, 23, 2 782.

[10] S. Krause, I. Streicher, P. Waltereit, L. Kirste, P. Brückner, S. Leone, IEEE Electron Device Letters 2023, 44, 1 17.

[11] J. Ligl, S. Leone, C. Manz, L. Kirste, P. Doering, T. Fuchs, M. Prescher, O. Ambacher, Journal of Applied Physics 2020, 127, 19 195704.

[12] C. G. Moe, J. Leathersich, D. Carlstrom, F. Bi, D. Kim, J. B. Shealy, physica status solidi (a) 2023, 220, 16 2200849.

[13] C. Manz, S. Leone, L. Kirste, J. Ligl, K. Frei, T. Fuchs, M. Prescher, P. Waltereit, M. A. Verheijen, A. Graff, M. Simon-Najasek, F. Altmann, M. Fiederle, O. Ambacher, Semiconductor Science and Technology 2021, 36, 3 034003.

[14] I. Streicher, S. Leone, L. Kirste, C. Manz, P. Straˇnák, M. Prescher, P. Waltereit, M. Mikulla, R. Quay, O. Ambacher, physica status solidi (RRL) – Rapid Research Letters 2023, 17, 2 2200387.

[15] A. Zukauskaite, G. Wingqvist, J. Palisaitis, J. Jensen, P. O. Persson, R. Matloub, P. Muralt, Y. Kim, J. Birch, L. Hultman, Journal of Applied Physics 2012, 111, 9 093527.

[16] X. Zhang, E. A. Stach, W. J. Meng, A. C. Meng, Nanoscale Horiz. 2023, 8 674.

[17] G. Schönweger, A. Petraru, M. R. Islam, N. Wolff, B. Haas, A. Hammud, C. Koch, L. Kienle, H. Kohlstedt, S. Fichtner, Advanced Functional Materials 32, 21 2109632.

[18] D. Wang, P. Wang, B. Wang, Z. Mi, Applied Physics Letters 2021, 119, 11 111902.

[19] G. Schönweger, M. R. Islam, N. Wolff, A. Petraru, L. Kienle, H. Kohlstedt, S. Fichtner, physica status solidi (RRL) Rapid Research Letters 2022, 2200312.

[20] G. Schönweger, N. Wolff, M. R. Islam, M. Gremmel, A. Petraru, L. Kienle, H. Kohlstedt, S. Fichtner, Advanced Science 10, 25 2302296.

[21] T. Sluka, A. K. Tagantsev, P. Bednyakov, N. Setter, Nature Communications 2013, 4, 1.

[22] E. A. Eliseev, A. N. Morozovska, G. S. Svechnikov, V. Gopalan, V. Y. Shur, Physical Review B 2011, 83, 23 235313.

[23] P. Wang, D. Wang, N. M. Vu, T. Chiang, J. T. Heron, Z. Mi, Applied Physics Letters 2021, 118, 22 223504.

[24] S. Yasuoka, T. Shimizu, A. Tateyama, M. Uehara, H. Yamada, M. Akiyama, Y. Hiranaga, Y. Cho, H. Funakubo, Journal of Applied Physics 2020, 128, 11 114103.

[25] N. Wolff, S. Fichtner, B. Haas, M. R. Islam, F. Niekiel, M. Kessel, O. Ambacher, C. Koch, B. Wagner, F. Lofink, L. Kienle, Journal of Applied Physics 2021, 129, 3 034103.

[26] F. Liu, R. Collazo, S. Mita, Z. Sitar, S. J. Pennycook, G. Duscher, Advanced Materials 2008, 20, 11 2162.

[27] S. Calderon, J. Hayden, S. M. Baksa, W. Tzou, S. Trolier-McKinstry, I. Dabo, J.-P. Maria, E. C. Dickey, Science 2023, 380, 6649 1034.

[28] Z. Liu, X. Wang, X. Ma, Y. Yang, D. Wu, Applied Physics Letters 2023, 122, 12 122901.

[29] S. Leone, J. Ligl, C. Manz, L. Kirste, T. Fuchs, H. Menner, M. Prescher, J. Wiegert, A. Žukauskaitˇe, R. Quay, O. Ambacher, physica status solidi (RRL) – Rapid Research Letters 2020, 14, 1 1900535.

[30] L. Jones, H. Yang, T. J. Pennycook, M. S. J. Marshall, S. V. Aert, N. D. Browning, M. R. Castell, P. D. Nellist, Advanced Structural and Chemical Imaging 2015, 1, 8.

[31] J. Barthel, Ultramicroscopy 2018, 193 1.

## A Supporting Figures

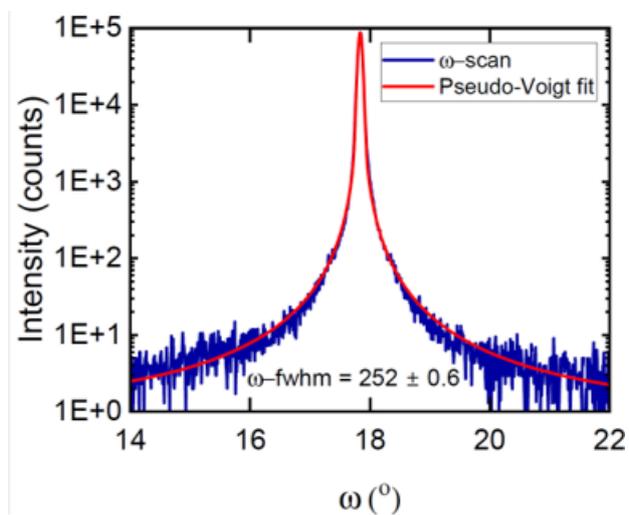

Figure A1: MOCVD grown Al0.85Sc0.15N : ω-scan of AlScN (0002) reflection with fwhm (in arcsec) from pseudo-voigt fitting.

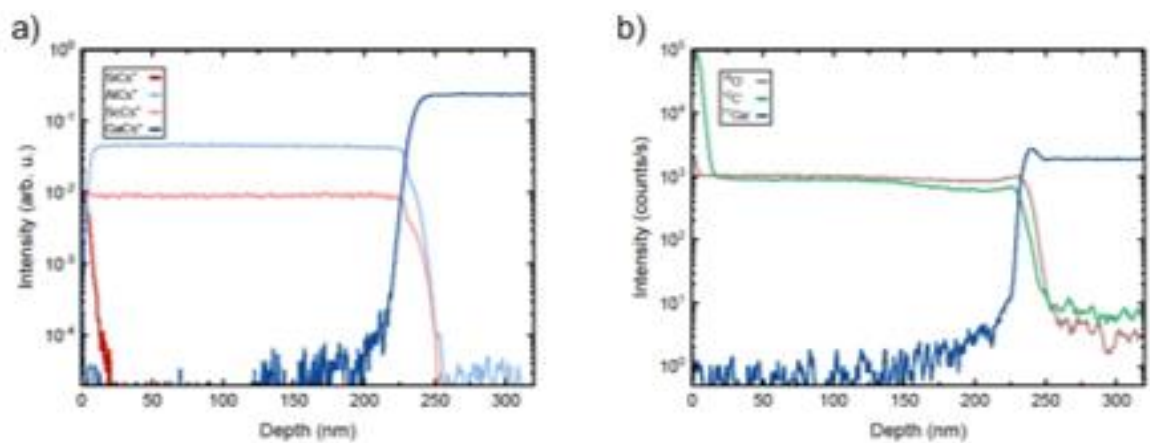

Figure A2: ToF-SIMS depth profiles of the MOCVD grown Al0.85Sc0.15N thin film on n-GaN. The signals obtained in (a) positive-ion detection mode were normalised to Cs2 + signal. (b) Negative-ion mode was used to obtain 12C- and 18O- intensity signals.

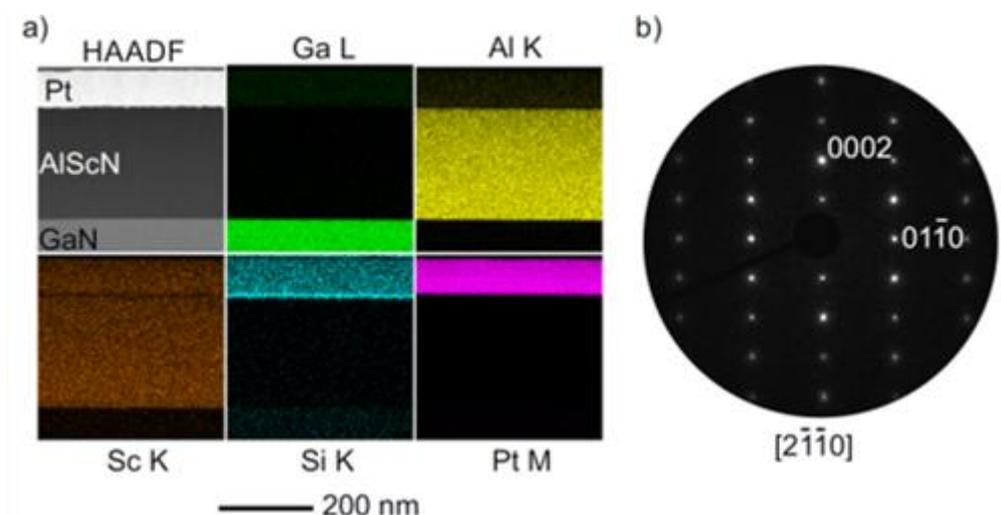

Figure A3: TEM analysis of the pristine Pt/SiN/Al0.85Sc0.15N/GaN heterostructure. a) HAADF-STEM micrograph of the heterostructure and the corresponding intensity maps recorded by EDS. b) Selected-area electron diffraction pattern recorded on the Al0.85Sc0.15N layer. The analysis demonstrates the single-crystalline structure of the film showing no chemical separation nor cubic inclusions. Sc and Si signals at the position of the Pt layer are due to the increased intensity background. The average concentration of Sc in the layer is x = 0.15.

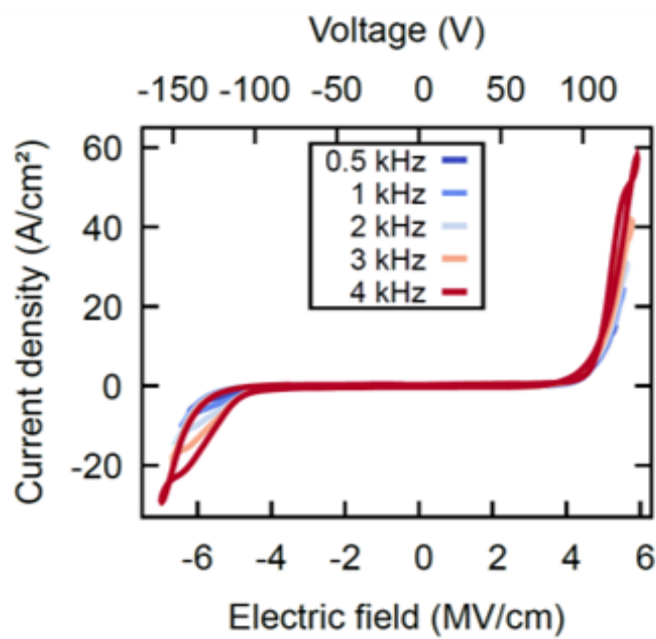

Figure A4: J − E loops in dependence of measurement frequency. The coercive field as well as the peak area increases with increasing frequencies, as expected for true ferroelectricity.

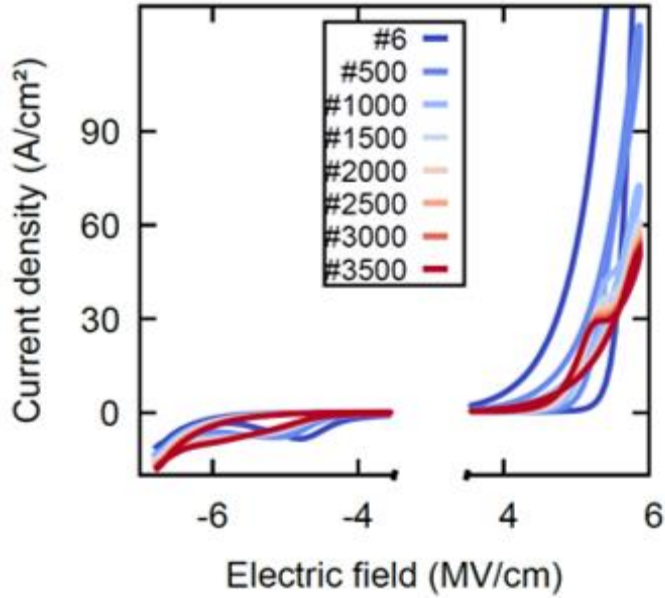

Figure A5: J − E loop evolution with cycling measured at 1.5 kHz. The huge hysteretic area appearing on as-grown pads at positive fields disappears after some tens of cycles. Further cycling shifts the coercive field towards more negative fields. On the positive side this favours the separation between leakage and ferroelectric displacement currents and the typical ferroelectric switching peak appears.

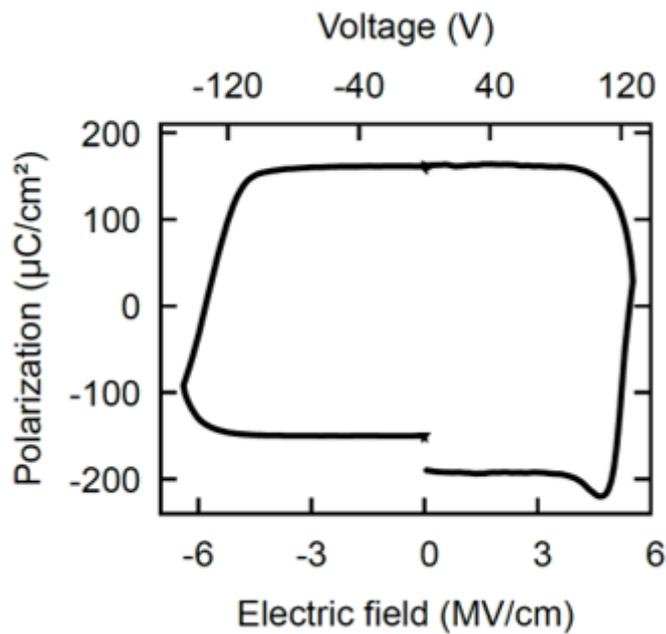

Figure A6: PUND corrected P−E loop measured at 1.5 kHz.

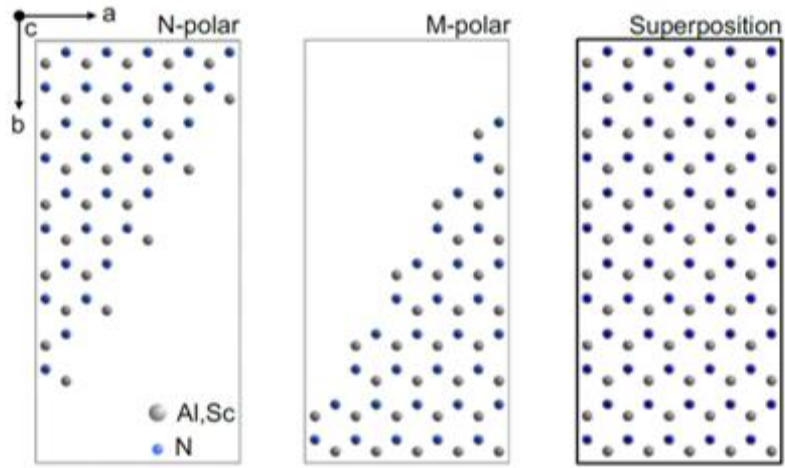

Figure A7: Supercell approach using two orthohexagonal cells of tapered N-polar and M-polar structure models viewed onto the c-plane. The combined superposition structure is used for simulation of the ABF scattering contrast along the b-c plane.